\documentclass{article}
\usepackage{latexsym}
\usepackage{subfigure}
\usepackage{dcolumn}
\usepackage{bm}
\usepackage{ulem}
\usepackage{color}
\usepackage[colorlinks,linkcolor=magenta,anchorcolor=cyan,citecolor=blue]{hyperref}
\usepackage{amsmath,amssymb,multicol,fancyhdr,geometry}
\usepackage{graphicx}
\usepackage{float}
\usepackage{mathrsfs}
\usepackage[numbers,sort&compress]{natbib}
\usepackage{verbatim}
\newcommand{\fig}[1]{Fig. \ref{#1}}
\def\be{\begin{equation}}
\def\ee{\end{equation}}
\def\ba{\begin{eqnarray}}
\def\ea{\end{eqnarray}}

\newcommand{\ppa}[2]{\left(\frac{\partial}{\partial #1}\right)^{#2}}

\newcommand{\eqn}{&=&}
\newcommand{\non}{\nonumber \\}
\newcommand{\eqs}[1]{Eqs. (\ref{#1})}
\newcommand{\eq}[1]{Eq. (\ref{#1})}
\newcommand{\meq}[1]{(\ref{#1})}

\newcommand{\bean}{\begin{eqnarray}}
\newcommand{\eean}{\end{eqnarray}}
\newcommand{\hsp}{\hspace{0.1mm}}

 \def\grad{\nabla}

\title{\textbf{Overcharging magnetized black holes at linear order and the weak cosmic censorship conjecture}}
\author{  Chengcheng Liu\footnote{Email:chengliu@mail.bnu.edu.cn}, Sijie Gao\footnote{Corresponding author. Email: sijie@bnu.edu.cn}\\
Department of Physics, Beijing Normal University,\\
Beijing 100875, China
}
\begin{document}
\maketitle
\begin{abstract}
Evidences have been found that the weak cosmic censorship conjecture could be violated if test particles  with charge and angular momentum are injected into a black hole. However, second-order corrections and fine-tunings on the particle's parameters are required in previous studies, indicating that self-force and radiative effects must be taken into account. In this paper, we first consider a magnetically charged particle falling into an extremal Bardeen black hole, which is regular (with no singularity) and has a magnetic monopole at the center. We then investigate a general class of magnetic black holes with or without singularities. In all the cases, we show that the test particle with magnetic charge could overcharge the black hole, causing possible violation of the weak cosmic censorship conjecture. In contrast to previous arguments in the literature, second-order corrections are not necessary in our analysis and the results are not sensitive to the particle's parameters. Our work indicates that the self-force effect, which is related to the second-order correction,  may not help rescue the weak cosmic censorship conjecture in our examples.

\end{abstract}

\section{Introduction}
In general relativity, a singularity could form in gravitational collapse of massive stars.  To avoid ``naked singularities'', Penrose \cite{c} introduced the weak cosmic censorship conjecture(WCCC) which states that singularities formed in gravitational collapse with physically reasonable matter must be covered by black hole horizons. The WCCC prevents singularities from being seen by distant observers. The general proof of WCCC is known to be extremely difficult. In a seminal work, Wald \cite{d}  proposed a gedanken experiment, showing that an extremal Kerr-Newman black hole cannot be destroyed by capturing a test particle. This work provided a strong support to WCCC. However, some possible counterexamples have been discussed in the past decades\cite{e1}-\cite{jiang}. By carefully selecting the parameters of the particle, it was found that  the black hole horizons could be destroyed in some cases. All the counterexamples require the second-order calculation in the particle's charge and energy, while in the analysis by Wald, only linear terms are involved. Consequently, the allowed window of the particle's parameters is very narrow and the self-force effect should be taken into account. Recently, Sorce and Wald \cite{f} gave a complete analysis on the second-order corrections and found that the WCCC is still valid for Kerr-Newman black holes.

Li an Bambi \cite{g}  discussed the possibility to destroy the horizon of a regular black hole, such as the rotating Bardeen black hole \cite{bardeen} and rotating Hayward black hole  \cite{i}. These black holes do not possess singularities. The source of the Bardeen black hole has been interpreted as a magnetic monopole \cite{ab1}-\cite{ab2}, which can be derived from  the nonlinear electrodynamics (NED). In \cite{g}, the test particle is neutral and then there is no electromagnetic interaction between the particle and the black hole. In this paper, we first send a test particle with magnetic charge into a Bardeen black hole to see if it can be overcharged. The first step is to derive the electromagnetic force exerted  on the particle. Because the Bardeen black hole is surrounded by the nonlinear electromagnetic field, we need  extend the well-known Lorentz force formula from  Maxwell's theory to NED. Starting from a general Lagrangian of nonlinear electrodynamics, we calculate the divergence of the stress-energy tensor and derive the Lorentz-like force exerted on the charged particle. Then we derive the conserved quantity along the charged particle moving in a stationary spacetime.
Application of these formulas shows that the horizon of the Bardeen black hole could be destroyed by such a particle. Differing from previous literature, we do not need take higher-order terms into account and the required parameter window is not necessarily narrow. We further investigate a general class of black holes with magnetic charges. With similar methods, we show that the black hole horizons could be destroyed under certain conditions. Particularly, we prove that some magnetic black holes with singularities could be overcharged, leading to a possible violation of the WCCC. Again, no higher-order terms are necessary in our treatment.
This is crucial because it implies that the self-force effect may not rescue the WCCC.

This paper is organized as follows. In section 2, we introduce the theory of NED and derive the extended Lorentz force exerted on a particle with magnetic charge. Then we derive the conserved quantity along the particle. In section 3, we send a particle with magnetic charge into the Bardeen black hole and demonstrate that the event horizon  could be destroyed and the allowed window of the particle's parameters is not necessarily narrow. In section 4, we investigate a general class of magnetized black holes with or without singularities and show that these black holes could be overcharged under certain conditions. Some concluding remarks are given in section 5.

\section{Nonlinear electrodynamics in curved spacetime}
In this section, we first introduce the field equations of NED in curved spacetime. Then we study the motion of a magnetically charged particle and derive the conserved quantities.

\subsection{Source-free Nonlinear electrodynamics}
It is well known that from the electromagnetic tensor $F_{ab}$ and its dual $\hspace{0.1mm}^*F_{ab}$, one can construct the two invariants
\bean
F\eqn F^{ab}F_{ab}\,, \label{novel} \\
Y\eqn\hspace{0.1mm}^*F^{ab}F_{ab}\,.
\eean
In the absence of magnetic monopoles,  $F_{ab}$ is a closed form and we can introduce the vector potential $A_a$ such that
\begin{gather}
F_{ab}=\grad_aA_b-\grad_bA_a\,,\label{yyz}
\end{gather}
which also gives
\begin{gather}
\grad_a\,^*F^{ab}=0\,.\label{yyzz}
\end{gather}

The NED Lagrangian is a function of $F$ and $Y$, denoted by $L_{NE}(F,Y)$. A general theory of sourceless nonlinear electromagnetic field in curved spacetime can be described by the action \cite{h}
\begin{gather}
S=\int d^4x\sqrt{-g}[R-L_{NE}(F,Y)]\,,\label{act}
\end{gather}
where $R$ is the scalar curvature.
For later convenience, we define the following  quantities
\bean
h\eqn \frac{\partial L_{NE}}{\partial F}\,,\\
 f\eqn\frac{\partial L_{NE}}{\partial Y} \,,  \\ \label{yyn}
G_{ab}\eqn hF_{ab}\,,   \label{ghf} \\
H_{ab}\eqn f\, \hspace{0.1mm}^*F_{ab}\,,\label{yyx}\\
L_{ab}\eqn G_{ab}+H_{ab}\,\label{yyc}.
\eean
Varying the action (\ref{act}) with respect to $A_a$ and $g_{ab}$ respectively yields the field equations
\begin{gather}
\grad_aL^{ab}=0\,,\label{dgz}
\end{gather}
and
\begin{gather}
R_{ab}-\frac{1}{2}Rg_{ab}=8\pi T_{ab}\,,\label{ein}
\end{gather}
where the stress-energy tensor $T_{ab}$ is given by
\begin{gather}
T_{ab}=\frac{1}{4\pi}(L_a\hspace{0.1mm}^c F_{bc}-\frac{1}{4}L_{NE}g_{ab})\,.\label{TT}
\end{gather}
\subsection{Derivation of the generalized Lorentz force}
To derive the electromagnetic force on a charged particle, we calculate
\begin{align}
\grad^aT_{ab}=&\frac{1}{4\pi}\grad^a(L_a\hspace{0.1mm}^c F_{bc}-\frac{1}{4}L_{NE}g_{ab})\nonumber\\
=&\frac{1}{4\pi}[F_{bc}\grad_aL^{ac}+(G^{ac}+H^{ac})\grad_aF_{bc}-\frac{1}{4}\grad_bL_{NE}]\nonumber\\
=&\frac{1}{4\pi}[F_{bc}\grad_aL^{ac}+(hF^{ac}+f \hspace{0.1mm}^*F^{ab})\grad_aF_{bc}-\frac{1}{4}(h\grad_bF+f\grad_bY)]\nonumber\\
=&\frac{1}{4\pi}[F_{bc}\grad_aL^{ac}+(hF^{ac}\grad_aF_{bc}-\frac{1}{4}h\grad_bF)+(f \hspace{0.1mm}^*F^{ab}\grad_aF_{bc}-\frac{1}{4}f\grad_bY)]\nonumber\\
=&\frac{1}{4\pi}F_{bc}\grad_aL^{ac}-\frac{3}{8\pi}(hF^{ac}+f\hspace{0.1mm}^*F^{ac})\grad_{[a}F_{cb]}\nonumber\\
=&\frac{1}{4\pi}F_{bc}\grad_aL^{ac}-\frac{1}{4\pi}(h\hspace{0.1mm}^*F_{bc}-fF_{bc})\grad_a\hspace{0.1mm}^*F^{ac}\nonumber\\
=&\frac{1}{4\pi}F_{bc}\grad_aL^{ac}-\frac{1}{4\pi}\hspace{0.1mm}(^*G_{bc}+\hspace{0.1mm}^*H_{bc})\grad_a\hspace{0.1mm}^*F^{ac}\nonumber\\
=&\frac{1}{4\pi}F_{bc}\grad_aL^{ac}-\frac{1}{4\pi}\hspace{0.1mm}^*L_{bc}\grad_a\hspace{0.1mm}^*F^{ac}\,.\label{sandu2}
\end{align}
In the source-free case, \eq{yyzz} and \eq{dgz} gives $\grad^a T_{ab}=0$, which is just the conservation law. However, if the particle carries electric and magnetic charge, we can define
the 4-electric and 4-magnetic current densities by
\bean
-4\pi J^c\eqn\grad_aL^{ac}\,,\label{JJ2}\\
-4\pi \hat J^c\eqn\grad_a\hspace{0.1mm}^*F^{ac}\,.\label{JJ1}
\eean
Then (\ref{sandu2}) becomes
\begin{gather}
\grad^aT_{ab}=-F_{bc}J^c+\,^*L_{bc}\hat J^c\,.\label{sandu}
\end{gather}
According to \eqs{ghf}-(\ref{yyc}),\meq{JJ2} and (\ref{JJ1}), we have
\begin{gather}
\grad_aJ^a=0\,,\,\grad_a \hat J^a=0\,,\label{Js1}
\end{gather}
which confirms the conservation of these currents.

For a point particle carrying electric charge $q$ and magnetic charge $g$, by integrating \eq{sandu} over the worldline, one can obtain
\begin{gather}
F_a=qF_{ab}U^b-g^*L_{ab}U^b\,,  \label{exlor}
\end{gather}
where $U^a$ is the four-velocity of the particle, i.e., the unit vector tangent to the worldline. We call this extended Lorentz force. If magnetic charges or monopoles are taken into account in Maxwell's theory, the Lorentz force takes exactly the same form as \eq{exlor} \cite{liang}. Authors \cite{h} derived the force on a particle with electric charge (but without magnetic charge) in NED, which is also consistent with \eq{exlor}.

\subsection{Conserved quantity of electric and magnetic charges}
In rest of this paper, we only consider nonlinear electromagnetic field with $Y=0$. So
\bean
L_{ab}=G_{ab} \,. \label{LG}
\eean
In the sourceless case, \eq{dgz} indicates that $\hsp^*G_{ab}$ is closed and then we can introduce the potential $\hat A_a$ such that
\begin{gather}
^*G_{ab}=\grad_a\hat{A}_b-\grad_b\hat{A}_a\,.\label{shineng2}
\end{gather}
According to \eq{exlor}, the equation of motion of a charged particle is determined by
\begin{gather}
mU^b\grad_bU_a=qF_{ab}U^b-g^*G_{ab}U^b\,.\label{Z6}
\end{gather}
If the spacetime possesses a Killing vector field $\xi^a$,
we can find a conserved quantity along the worldline. For this purpose,
we calculate
\bean
&&U^b\grad_b(m\xi^aU_a)\non
\eqn m\xi^a U^b\grad_bU_a \non
\eqn qU^b[\xi^a\grad_aA_b-\grad_b(\xi^aA_a)+A_a\grad_b\xi^a]
-gU^b[\xi^a\grad_a\hat{A}_b-\grad_b(\xi^a\hat A_a)+\hat{A}_a\grad_b\xi^a]\,,\label{oppor11}
\eean
where in the first step, we have used the antisymmetry of $\grad_a\xi_b$ and in the second step, \eqs{yyz}, \meq{shineng2} and \meq{Z6} have been used.

By rearranging \eq{oppor11}, we find
\begin{align}
&U^b\grad_b(m\xi^aU_a+q\xi^aA_a-g\xi^a\hat{A}_a)\nonumber\\
&=qU^b(\xi^a\grad_aA_b+A_a\grad_b\xi^a)-gU^b(\xi^a\grad_a\hat{A}_b+\hat{A}_a\grad_b\xi^a)\nonumber\\
&=qU_b(\xi^a\grad_aA^b-A_a\grad^a\xi^b)-gU_b(\xi^a\grad_a\hat{A}^b-\hat{A}_a\grad^a\xi^b)\nonumber\\
&=qU_b(\mathcal{L}_\xi A^b)-gU_b(\mathcal{L}_\xi \hat{A}^b)\nonumber\\
&=0\,.
\end{align}
Therefore, the conserved quantity associated with $\xi^a$ is
\begin{gather}
P=m\xi^aU_a+q\xi^aA_a-g\xi^a\hat{A}_a\,\label{shou}.
\end{gather}
This quantity will play an important role in the following calculation.

\section{Destroying the event horizon of a Bardeen black hole}\label{sec-bardeen}
In this section, we shall check whether a magnetically charged particle can destroy the event horizon of a Bardeen black hole.

The Bardeen model can described by the line element\cite{bardeen}
\begin{gather}
ds^2=-\left[1-\frac{2Mr^2}{(r^2+\hat Q^2)^{\frac{3}{2}}}\right]dt^2+\left[1-\frac{2Mr^2}{(r^2+\hat Q^2)^{\frac{3}{2}}}\right]^{-1}dr^2+r^2d\Omega^2\,\label{dsB}.
\end{gather}
According to \cite{ab2}, the solution is generated from a nonlinear electrodynamic field with Lagrangian
\bean
L_{NE}(F)=\frac{12M}{|\hat Q^3|}\left(\frac{\sqrt{\hat Q^2F}}{\sqrt{2}+\sqrt{\hat Q^2 F}}\right)^{\frac{5}{2}}\,,
\eean
where
\bean
F_{ab}=\hat Q sin\theta(d\theta_ad\phi_b-d\theta_bd\phi_a)\,.\label{yya}
\eean
By virtue of \eqs{JJ2} and \meq{JJ1}, we can obtain the total electric charge and magnetic charge of the black hole
\bean
Q_e\eqn\frac{1}{4\pi}\int_s\,^*G_{ab}\,,\label{defq}\\
Q_m\eqn\frac{1}{4\pi}\int_s F_{ab}\,,\label{defqh}
\eean
where S is any two-sphere surrounding the black hole.

By employing \eqs{defq} and \meq{defqh}, we see that the Bardeen black hole possesses no electric charge and  $\hat Q$ is just its magnetic charge. So the source of the Bardeen solution is a magnetic monopole located at the center $r=0$ \cite{ab1,ab2}. The black hole is \textit{regular} because it has no singularity.

By definition, one can calculate
\begin{gather}
^*G_{ab}=h^*F_{ab}=\frac{15\hat Q M r^4}{2(\hat Q^2+r^2)^{7/2}}(dt_adr_b-dr_adt_b)\,,
\end{gather}
and the corresponding potential
\bean
\hat{A}_a= -\frac{3M}{2\hat Q}\biggl[\frac{r^5}{(\hat Q^2+r^2)^{\frac{5}{2}}}-1\biggr](dt)_a\,,\label{g2}
\eean
where the gauge has been fixed by the condition $\hat A_a \rightarrow 0$  at infinity.

From \eq{dsB}, we see that the horizon of the black hole can be found by solving
\bean
1-\frac{2Mr^2}{(r^2+\hat Q^2)^{\frac{3}{2}}}=0\,.
\eean
This is a cubic equation of $r^2$.
For
\bean
\hat Q^2<\frac{16}{27}M^2\,,
\eean
the black hole possesses two horizons.

 For
\bean
\hat Q^2=\frac{16}{27}M^2  \,,\label{ji2}
\eean
there is only one horizon and the black hole is called \textit{extremal}.

For
\bean
\hat Q^2>\frac{16}{27}M^2  \,, \label{qbm}
\eean
there is no horizon.

Consider a particle with mass $m$ and magnetic charge $g$ moving outside the extremal Bardeen black hole. If the particle can enter the black hole and the parameters of the resulting spacetime satisfy \eq{qbm}, the black hole is called \textit{overcharged}  and its horizon could disappear. The conserved quantity associated with the Killing vector field $\ppa{t}{a}$ is the energy, denoted by $E$. By considering the radial motion, \eq{shou} yields

\begin{align}
E=&-\biggl(mg_{tt}\frac{dt}{d\tau}+qA_t-g\hat A_t\biggr)\nonumber\\
=&-\frac{3Mg}{2\hat Q}\biggl[\frac{r^5}{(\hat Q^2+r^2)^{\frac{5}{2}}}-1\biggr]+m\frac{dt}{d\tau}\biggl[1-\frac{2Mr^2}{(\hat Q^2+r^2)^{\frac{3}{2}}}\biggr]\,.\label{lkk}
\end{align}

The four-velocity takes the form
\bean
 U^a=\ppa{\tau}{a}=\dot t\ppa{t}{a}+\dot r \ppa{r}{a}\,,
\eean
  where $\tau$ is the proper time of the particle and the dot denotes  derivative with respect to $\tau$. Thus, the
 normalization condition is written as
\begin{align}
-1=&g_{ab}U^aU^b\nonumber\\
=&-\biggl[1-\frac{2Mr^2}{(r^2+\hat Q^2)^{\frac{3}{2}}}\biggr]\dot t^2+\biggl[1-\frac{2Mr^2}{(r^2+\hat Q^2)^{\frac{3}{2}}}\biggr]^{-1}\dot r^2\,,\label{lkkk}
\end{align}
which gives
\bean
\dot t=\pm \left[1-\frac{2Mr^2}{(r^2+\hat Q^2)^{3/2}}\right]^{-1}\sqrt{\dot r^2+\left[1-\frac{2Mr^2}{(r^2+\hat Q^2)^{3/2}}\right]} \label{tdot}
\eean
Using the fact that $U^a$ is future-directed, one can show that $\dot t>0$. So we should take the positive sign in \eq{tdot}. Substituting \eq{tdot} into \eq{lkk}, we have
\begin{align}
E=\frac{3gM}{2\hat Q}\biggl[1-\frac{r^5}{(\hat Q^2+r^2)^{\frac{5}{2}}}\biggr]+m\sqrt{\dot r^2+\biggl[1-\frac{2M r^2}{(\hat Q^2+r^2)^{\frac{3}{2}}}\biggr]}\,.\label{lkl}
\end{align}

We shall be interested in the extremal case, where the charge and mass are related by \eq{ji2}. It is easy to find that the horizon of the black hole is located at
\begin{gather}
r_h=\sqrt{\frac{32}{27}}M\label{hor}\,.
\end{gather}
Substitution of \eqs{ji2} and \meq{hor} into \eq{lkl} yields
\bean
E\geq \frac{3gM}{2\hat Q}\left[1-\frac{r^5}{(\hat Q^2+r_h^2)^{\frac{5}{2}}}\right]=\frac{9\sqrt{3}-4\sqrt{2}}{8}\frac{\hat Q}{|\hat Q|}g \,. \label{mime}
\eean
This just gives the minimum energy that allows a particle to reach the horizon of the black hole.

On the other hand, \eq{qbm} suggest that
\begin{align}
(\hat Q+g)^2>\frac{16}{27}(M+E)^2 \label{n12a}
\end{align}
must hold if the particle can destroy the horizon.
Since $g \ll \hat Q$ and $E\ll  M$, by keeping the linear terms, we have
\begin{align}
\frac{\hat Q}{|\hat Q|}g>\frac{4}{\sqrt {27}} E\,.\label{n12}
\end{align}
Without loss of generality, we assume $\hat Q>0$ and $g>0$. Thus, Combining \eqs{mime} and \meq{n12}, we obtain
\begin{gather}
\frac{8}{9\sqrt{3}-4\sqrt{2}}>\frac{g}{E}>\frac{4}{\sqrt{27}}\label{n5},
\end{gather}
or
\bean
0.806>\frac{g}{E}>0.770\,.
\eean
If the parameters of the particle fall into this interval, the particle could enter the horizon and destroy the horizon. Note that to obtain the above result, we only took the terms linear to $g$ and $E$. This is a crucial difference from previous examples.

Next, we check whether a particle released from infinity can overcharge the blackhole. For this purpose,
 we define the effective potential
\begin{gather}
V_{eff}=-\dot r^2\,.\label{uiu}
\end{gather}
By solving \eq{lkl}, we can write the effective potential of the extremal black hole as
\begin{gather}
V_{eff}=1-\frac{3\sqrt{3}x^2}{2(1+x^2)^{\frac{3}{2}}}-\frac{1}{m^2}\biggl\{E-\frac{9\sqrt{3}g}{8}\biggl[1-\biggl(\frac{1}{x^2}+1\biggr)^{-\frac{5}{2}} \biggr]\biggr\}^2\,,    \label{LO}
\end{gather}
where $x=\frac{r}{\hat Q}$. Suppose that the particle stays still at infinity and then is released, which means $E=m$.  Now \eq{LO} reduces to
\begin{align}
V_{eff}=1-\frac{3\sqrt{3}x^2}{2(1+x)^{\frac{3}{2}}}-\biggl\{1-\frac{9\sqrt{3}g}{8m}\biggl[1-\biggl(\frac{1}{x^2}+1\biggr)^{-\frac{5}{2}} \biggr]\biggr\}^2\,.\label{n555}
\end{align}

Taking into account \eq{n5}, it is easy to check $V_{eff}$ in \eq{n555} increases with  $g/m$. So by substituting   the upper bound in \eq{n5} into the right-hand side of \eq{n555}, we have
\begin{align}
V_{eff}\leq
1-\frac{3\sqrt{3}x^2}{2(1+x^2)^{3/2}}-\left\{1-\frac{27[1-x^5 (1+x^2)^{-5/2}]}{27-4\sqrt{6}}\right\}^2\,.\label{LK}
\end{align}
One can check that the right-hand side of (\ref{LK}) is always negative outside the horizon ( $x>\sqrt{2}$). This means that the particle released from infinity can go all the way to the horizon and then overcharge the Bardeen black hole.

\section{General magnetized black holes and the WCCC}
In this section, we will demonstrate the event horizons of a general class of magnetized black holes proposed by \cite{a} can be destroyed by test particles. Particularly, some of them contain singularities, leading to possible violation of the WCCC.

\subsection{Magnetized black holes}
A static and spherically symmetric metric is described by the line element
\begin{gather}
ds^2=-f(r)dt^2+\frac{1}{f(r)}dr^2+r^2d\Omega^2  \,\label{ds}.
\end{gather}
Introduce the mass function $m(r)$ by
\begin{gather}
f(r)=1-\frac{2m(r)}{r}\,.\label{6788}
\end{gather}
The black hole horizon is located at $r=r_h$ satisfying
\bean
r_h=2m(r_h)\,.
\eean
If the black hole is coupled to nonlinear electrodynamics, we can substitute \eqs{TT} and \meq{LG} into Einstein's equation (\ref{ein}) and find the nonzero components of $F_{ab}$:
\bean
F_{tr}\eqn-F_{rt}=\pm\sqrt{\frac{1}{2h}\biggr[\frac{2m'(r)}{r^2}-\frac{1}{2}L_{NE}(F)\biggl]}\,,\label{Ftr}\\
F_{\theta \phi}\eqn-F_{\phi \theta}=\pm r^2\sin\theta\sqrt{\frac{1}{2h}\left(\frac{1}{2}L_{NE}(F)-\frac{m''(r)}{r}\right)} \,.\label{JJ9}
\eean

By performing the integrations in \eqs{defq} and \meq{defqh} over a two-sphere $S$, we obtain the electric  charge and magnetic charge of the black hole

\bean
Q\eqn \frac{1}{8\pi}\int_s\,G^{cd}\epsilon_{cdab}=-\frac{r^2}{4\pi}\int_s\,hF_{tr}sin\theta d\theta d\phi=-r^2h F_{tr}\,,\label{JJ5}\\
\hat Q\eqn\frac{1}{4\pi}\int_s F_{\theta\phi}d\theta d\phi=\frac{F_{\theta\phi}}{sin\theta}\,.\label{JJ6}
\eean
Therefore,
\begin{gather}
F_{ab}=-\frac{Q}{r^2h}(dt_adr_b-dr_adt_b)+\hat{Q}sin\theta(d\theta_ad\phi_b-d\theta_bd\phi_a)\,,\label{oppor155}
\end{gather}
and
\begin{gather}
^*G_{ab}=\frac{h\hat Q}{r^2}(dt_adr_b-dr_adt_b)+Qsin\theta(d\theta_ad\phi_b-d\theta_bd\phi_a)\,.\label{oppor15}
\end{gather}
Combining (\ref{yyz})(\ref{shineng2}) and (\ref{oppor155})(\ref{oppor15}), we get the potentials
\begin{gather}
A_a=Q\int\frac{dr}{r^2h'(F)}(dt)_a-\hat Qcos\theta(d\phi)_a\,\label{f1},\\
\hat A_a=-\hat Q\int\frac{h dr}{r^2}(dt)_a-Qcos\theta(d\phi)_a\,\label{f2}.
\end{gather}
From now on, we only consider  magnetized black holes, i.e.,
\bean
 Q=0\,, \label{Q0}
\eean
and
\bean
F_{tr}=0\,. \label{ftrz}
\eean
Thus,
\begin{gather}
F=2F_{\theta\phi}F^{\theta\phi}=2\frac{\hat Q^2}{r^4}\,.\label{Ftt}
\end{gather}
Combining \eqs{Ftr} and (\ref{ftrz}), we have
\begin{gather}
L_{NE}(F)=4\frac{m'(r)}{r^2}\,.\label{678}
\end{gather}
Substituting \eqs{Q0}(\ref{Ftt})(\ref{678}) into (\ref{f1})(\ref{f2})
\begin{align}
A_a&=-\hat Qcos\theta (d\phi)_a\,,\label{mma}\\
\hat A_a&=-\hat Q\int\frac{1}{r^2}\frac{dL_{NE}}{dF} dr(dt)_a\nonumber\\
&=-\hat Q\int\frac{1}{r^2}\frac{dr}{dF}dL_{NE}(dt)_a\nonumber\\
&=\frac{1}{8\hat Q}\int r^3dL_{NE}(dt)_a\nonumber\\
&=\frac{1}{2\hat Q}\int (rm''-2m')dr (dt)_a   \nonumber\\
&=\frac{1}{2\hat Q}[-3m(r)+rm'(r)+3m(r)|_{r=\infty}](dt)_a\,.\label{mmb}
\end{align}
In the last step, we have assumed that $rm'(r)$ vanishes at infinity and the integration constant has been chosen such that $\hat A_a=0$ as $r\rightarrow \infty$.

\subsection{Energy of a test particle}
We only consider radial movement of a test particle with mass $m$ and magnetic charge $g$. Substituting \eq{mma} and (\ref{mmb}) into (\ref{shou}) and taking into account \eq{ds}, we obtian the conserved energy
\begin{align}
E=&-\left(mg_{tt}\dot t-g\hat A_t\right)\nonumber\\
=&mf(r)\dot t+\frac{g}{2\hat Q}\left[-3m(r)+rm'(r)+3m(r)|_{r=\infty}\right]\,.\label{ddd}
\end{align}
Again, the normalization condition yields
\begin{align}
-1=&g_{ab}U^aU^b=-f(r)\dot t^2+f(r)^{-1}\dot r^2\label{dddd}\,.
\end{align}
Combining (\ref{ddd}) and (\ref{dddd}), we have
\begin{gather}
\dot r^2=\frac{1}{m^2}\left\{E-\frac{g}{2\hat Q}[-3m(r)+rm'(r)+3m(r)|_{r=\infty}]\right\}^2-1+\frac{2m(r)}{r}\label{345}\,.
\end{gather}
By solving \eq{345}, we have
\begin{align}
E=\frac{g}{2\hat Q}[-3m(r)+rm'(r)+3m(r)|_{r=\infty}]+m\sqrt{\left(\frac{dr}{d\tau}\right)^2+\left[1-\frac{2m(r)}{r}\right]}\label{nengli}\,.
\end{align}
At the horizon $r=r_h$, an extremal black hole satisfies
\bean
f(r_h)\eqn0\,,\\
f'(r_h)\eqn0\,.\label{rtr}
\eean
Therefore, \eq{6788} gives
\bean
2m(r_h)\eqn r_h\,, \label{mrh} \\
m'(r_h)\eqn\frac{1}{2}\,.\label{gggg}
\eean
Substituting \eq{mrh} and (\ref{gggg}) into \eq{nengli}, we find the constraint for the particle to reach the horizon:
\bean
E\geq\frac{g}{2\hat Q}[-r_h+3m(r)|_{r=\infty}]\,.\label{zuixiao}
\eean

\subsection{Overcharging some general magnetized black holes}
A generic class of magnetized black hole solutions have been proposed in \cite{a}\cite{b} by taking the mass function in the form
\begin{gather}
m(r)=\frac{Mr^{\mu}}{(r^\nu+\hat Q^\nu)^{\frac{\mu}{\nu}}}\,.\label{4444}
\end{gather}
The corresponding Lagrangian is
\begin{gather}
L_{NE}(F)=\frac{4\mu}{\alpha}\frac{(\alpha F)^{\frac{\nu+3}{4}}}{[1+(\alpha F)^{\nu/4}]^{1+\frac{\mu}{\nu}}}\,.\label{oppor17}
\end{gather}
 Here $\mu> 0$ and $\nu> 0$ are dimensionless constants and the value of $\nu$ characterizes the strength of the nonlinear electromagnetic field in the weak field limit. \eq{4444} covers several well-known black hole solutions. For example, $\nu=1$, $\nu=2$ and $\nu=3$ correspond to the Maxwellian solution in the weak field regime, Bardeen-like solutions, and Hayward-like solutions, respectively.
  For $\mu\geq 3$,  the black hole is regular. For $0<\mu< 3$, the black hole possesses a singularity. More properties  have been studied in  \cite{a},\cite{6}-\cite{tosh}.

\begin{figure}[H]
\centering
\includegraphics[width=0.5\textwidth]{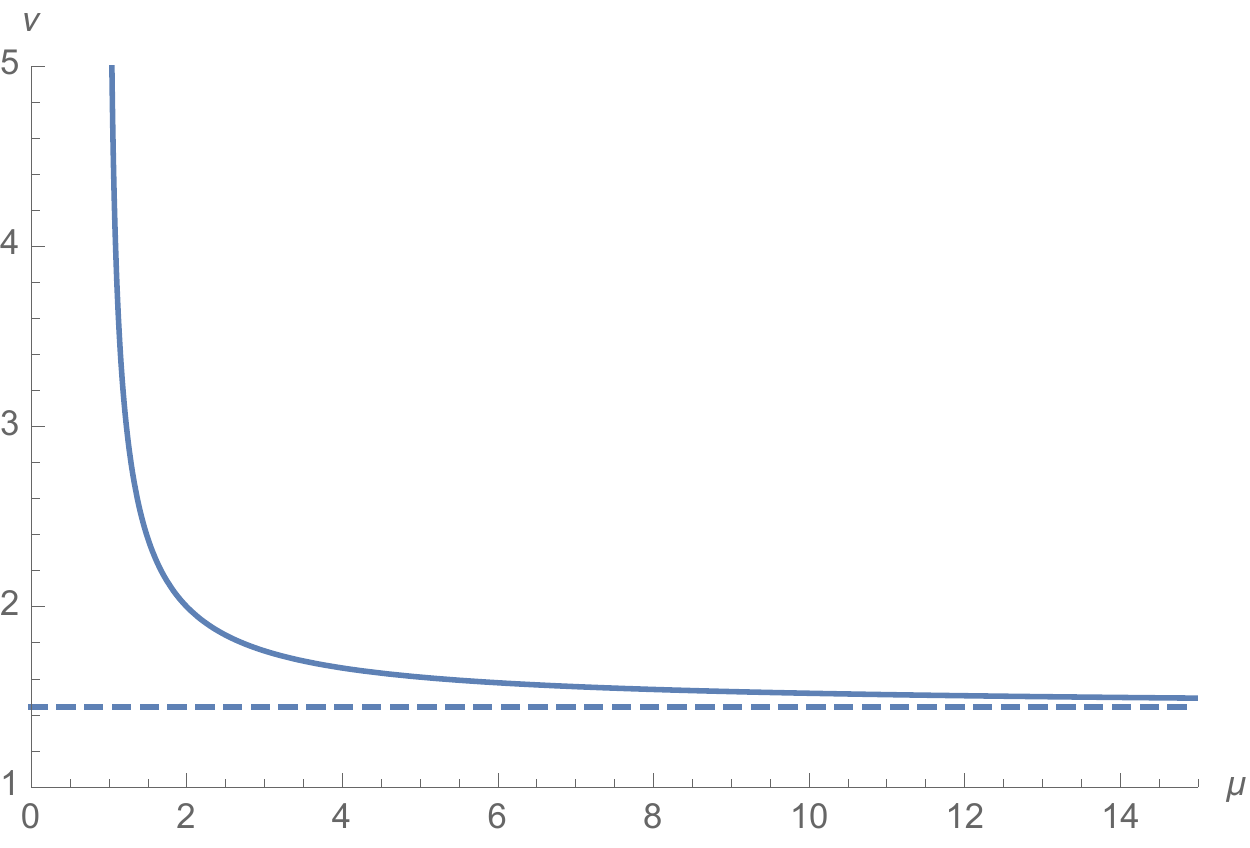}
\caption{The plot of $\mu-\nu $. Above the solid line, the horizon could be destroyed.}\label{ABC}
\end{figure}

One can show that $M$ is the gravitational mass and $\hat Q$ is the magnetic charge \cite{b,tosh}. We still consider an extremal black hole and assume $\mu>1$.
From \eqs{mrh},(\ref{gggg}) and (\ref{4444}), it is not difficult to find
\bean
\hat Q^\nu\eqn (2M)^\nu\frac{(\mu-1)^{\mu-1}}{\mu^\mu}\label{ji}\,,\\
r_h\eqn2M\left(1-\frac{1}{\mu}\right)^{\frac{\mu}{\nu}}\,,\label{jx} \\
M\eqn m(r)|_{r=\infty}\label{zsx}\,.
\eean
 Without loss of  generality, we assume $\hat Q >0$. Consider a test particle with energy $E$ and  magnetic charge $g>0$. Substituting (\ref{ji})(\ref{zsx}) into (\ref{zuixiao}), we find the condition for the particle to reach the horizon:
\begin{gather}
E> \frac{g}{4}\left[\frac{\mu^\mu}{(\mu-1)^{\mu-1}}\right]^{\frac{1}{\nu}}\left[3-2\left(1-\frac{1}{\mu}\right)^{\frac{\mu}{\nu}}\right]\label{n2}\,.
\end{gather}
On the other hand, in order to overcharge the black hole, the following inequality must be satisfied
\begin{align}
(\hat Q+g)^\nu>(2M+2E)^\nu\frac{(\mu-1)^{\mu-1}}{\mu^\mu}\,. \label{n1}
\end{align}
where \eq{ji} has been used. Since $g\ll \hat Q$ and $E\ll M$, by taking the linear orders, \eq{n1} becomes
\begin{align}
E<\frac{g}{2}\left[\frac{\mu^\mu}{(\mu-1)^{\mu-1}}\right]^{\frac{1}{\nu}}\label{fgf}\,.
\end{align}
Puting \eq{n2} and (\ref{fgf}) together, we obtain
\begin{gather}
\frac{g}{4}\left[\frac{\mu^\mu}{(\mu-1)^{\mu-1}}\right]^{\frac{1}{\nu}}\left[3-2
\left(1-\frac{1}{\mu}\right)^{\frac{\mu}{\nu}}\right]< E<\frac{g}{2}\left[\frac{\mu^\mu}{(\mu-1)^{\mu-1}}\right]^{\frac{1}{\nu}}\,.\label{n7}
\end{gather}
The existence of $E$ in \eq{n7} requires
\begin{gather}
2^{\nu}>\biggl(\frac{\mu}{\mu-1}\biggr)^\mu\,. \label{vbu}
\end{gather}
As shown in \fig{ABC}, \eq{vbu} is satisfied in the region above the solid line.

For $\nu=1$ and in the weak field limit $F\ll 1$, the solution  reduces to the Maxwellian solution, which is obviously below the solid line. This is consistent with the previous conclusion that black holes associated with Maxwell's theory cannot be overcharged by test particles.

When $\mu=3$ and $\nu=2$,  the solution is just the Bardeen black hole, as we have discussed.

When $\mu=3$ and $\nu=3$, we recover the Hayward black hole \cite{i}, which is also regular. \eq{n7} becomes
\begin{gather}
\frac{5}{4\sqrt[3]{4}}g< E< \frac{3}{2\sqrt[3]{4}}g\,,
\end{gather}
or
\bean
0.79g<E<0.94 g
\eean

It is important to consider the case $\mu=2$ and $\nu=3$ because the solution describes a black hole with singularity\cite{a}. It follows from \eq{n7} that
\begin{gather}
0.69 g\leq E<0.79g\,. \label{lee}
\end{gather}
Therefore, the black hole could be destroyed and the singularity becomes naked. One can check that this spacetime satisfies the weak energy condition. Furthermore, from \eq{lee} we calculate
\bean
\frac{\Delta E}{\bar E}=\frac{0.79g-0.69 g}{(0.79g+0.69 g)/2}=0.14\,.
\eean
This value is much larger than that in other cases. For example, the corresponding value for an extremal Kerr-Newman black hole is $ \frac{\Delta E}{\bar E}\approx 10^{-3}$  \cite{e11}. This is because higher order terms considered in previous literature are much smaller than the linear order terms in our analysis.

Similar to the treatment at the end of section \ref{sec-bardeen}, we find that the effective potential satisfies
\begin{align}
V_{eff}<-\left\{1-\left[3-2\left(1-\frac{1}{\mu}\right)^{\frac{\mu}{\nu}}\right]^{-1}
\left[3+\frac{x^\mu(-3x^\nu+\mu-3)}{(1+x^\nu)^{1+\mu/\nu}}\right]\right\}^2
+1-\left[\frac{\mu^\mu}{(\mu-1)^{\mu-1}}\right]^{1/\nu}
\frac{x^{\mu-1}}{(x^\nu+1)^{\mu/\nu}}\label{huawei}\,,
\end{align}
where $x=r/\hat Q$.  For $\mu=2$ and $\nu=3$, we have
\begin{gather}
V_{eff}<1-\frac{\sqrt[3]{4}x}{(1+x^3)^{2/3}}-\left\{-1+\frac{1}{3-\sqrt[3]{2}}
[3-x^2(1+3x^3)(1+x^3)^{-5/3}]\right\}\,.
\end{gather}
It's easy to check that this formula is always negative outside the horizon($x>1$). So the test particle can be released from infinity and overcharge this extremal black hole.

\section{Conclusions} \label{sec-con}
In this paper, we first derived the Lorentz-like force on a magnetically charged particle in the NED theory. We then show that such a particle could enter the Bardeen black hole and make its horizon disappear. Furthermore, we explored a general class of magnetically charged black holes. In particular, we have shown that a particle with magnetic charge could overcharge a black hole with singularity, leading to a possible violation of the WCCC. In contrast to previous gedanken experiments, our results do not require higher order terms and fine-tunings on the particle's parameters. This indicates that the second-order effects, like the self-force, may not rescue the WCCC in this case.  Although no experimental evidence for the existence of magnetic charges or monopoles  has been found yet, our work could shed some new light on the research of WCCC.

\section*{Acknowledgements}
 This research was supported by NSFC Grants No. 11775022 and 11873044.


\begin{thebibliography}{99}
\bibitem{c} R. Penrose, Revistas del Nuovo Cimento \textbf {1}, 252 (1969).
\bibitem{d} R. Wald, Annals of Physics \textbf {82}, 548 (1974).




\bibitem{e1} V.E.Hubeny, Phys. Rev. D \textbf {59}, 064013 (1999).
\bibitem{e2} S.Hod, Phys. Rev. D \textbf {66}, 024016 (2002).

\bibitem{e5} G.Chirco, S.Liberati and T.P.Sotiriou, Phys. Rev. D \textbf {82}, 104015 (2010).
\bibitem{e6} B.Gwak and B.H.Lee, Phys. Rev. D \textbf {84}, 084049 (2011).
\bibitem{e7} B.Gwak and B.H.Lee, Class. Quant. Grav. \textbf {29}, 175011 (2012).
\bibitem{e8} G.Z.Toth, Gen. Relativ. Gravit. \textbf {44,} 2019-2035 (2010).
\bibitem{e9} M.Bouhmadi-Lopez, V.Cardoso, A.Nerozzi, J.V. Rocha, Phys. Rev D \textbf {81}, 084051 (2010).
\bibitem{e10} J.V. Rocha and V.Cardoso, Phys.Rev. D \textbf {83,} 104037, (2011).
\bibitem{e4}T.Jacobson and T.P.Sotiriou, J. Phys.: Conf. Ser. \textbf {222}, 012041 (2010).
\bibitem{e3} T.Jacobson and T.P.Sotiriou, Phys. Rev. Lett \textbf {103}, 141101 (2009).
\bibitem{e11} S. Gao and Y. Zhang, Phys. Rev. D \textbf {87}, 044028 (2013).
\bibitem{e12} M. Colleoni and L. Barack, Phys. Rev. D \textbf {91}, 104024 (2015).
\bibitem{e13} M. Colleoni, L. Barack, A. G. Shah, and M. van de Meent, Phys. Rev. D \textbf {92}, 084044 (2015).
\bibitem{liuyuxiao} B.Liang, S.Wei, Y. Liu, Mod. Phys. Lett. A {\bf 34,}1950037 (2019).
\bibitem{nb} B.Ning, B.Chen, F.L.Lin, Phys.Rev. D {\bf 100,} 044043 (2019).
\bibitem{jiang} J.Jiang, X.Liu and M.Zhang, Phys.Rev.D \textbf{100,} 084059 (2019).

\bibitem{f} J. Sorce and R. M. Wald, Phys. Rev. D \textbf{96}, 104014 (2017).
\bibitem{g} Z. Li and C. Bambi, Phys. Rev. D \textbf {87}, 124022 (2013).
\bibitem{bardeen} J. M. Bardeen, in \textit {Conference Proceedings of GR5} (Tbilisi, USSR, 1968), p. 174.
\bibitem{i}S. A. Hayward, Phys. Rev. Lett. \textbf {96}, 031103 (2006).
\bibitem{ab1} E. Ay\'{o}n-Beato and A. Garc\'{i}a, Phys. Rev. Lett. \textbf {80}, 5056 (1998).
\bibitem{ab2} E. Ay\'{o}n-Beato and A. Garc\'{i}a, Phys.Lett. B \textbf {493}, 149 (2000).
\bibitem{h}H. Li, X. Yang and J. Wang, arXiv:1906.11702 [astro-ph.HE].
\bibitem{liang} C.B. Liang and B.Zhou, \textit {Fundamentals of Differential Geometry and General Relativity}, Vol II (Science Press, Beijing 2009). p. 78.
\bibitem{a} Z.-Y. Fan and X. Wang, Phys. Rev. D \textbf {94,}124027 (2016).
\bibitem{b} B. Toshmatov, Z. Stuchlik, B. Ahmedov and D. Malafarina, Phys. Rev. D \textbf {99}, no. 6, 064043 (2019).
\bibitem{6} K.A.Bronnikov, Phys.Rev. D \textbf {96,} 128501 (2017).
\bibitem{tosh} B.Toshmatov, Z. Stuchl\'{i}k, and B. Ahmedov, Phys. Rev. D \textbf {98}, 028501 (2018).
\end{thebibliography}
\end{document}